# Far-reaching consequences of trait preferences for animal social network structure and function

Josefine Bohr Brask[1,2,3,*], Andreas Koher[3], Darren P. Croft[4], Sune Lehmann[3]

*1. Copenhagen Center for Social Data Science (SODAS), University of Copenhagen, Denmark*

*2. Section for Ecology and Evolution, Department of Biology, University of Copenhagen, Denmark*

*3. DTU Compute – Department of Applied Mathematics and Computer Science, Technical University of Denmark (DTU), Denmark*

*4. Centre for Research in Animal Behaviour, Faculty of Health and Life Sciences, University of Exeter, United Kingdom*

**Corresponding author. Address: Øster Farimagsgade 5A, Building 1, 2nd floor, 1353 Copenhagen K, Denmark. Email: jbb@sodas.ku.dk; bohrbrask@gmail.com*

## Abstract

Social network structures play an important role in the lives of animals by affecting individual fitness and the spread of disease and information. Nevertheless, we still lack a good understanding of how these structures emerge from the behavior of individuals. Generative network models provide a powerful approach that can help close this gap. Empirical research has shown that trait-based social preferences (preferences for social partners with certain trait values, such as sex, body size, relatedness etc.) play a key role in the formation of social networks across species. Currently, however, we lack a good understanding of how such preferences affect network properties. In this study: 1) we develop a general and flexible generative network model that can create artificial (simulated) networks where social connection is affected by trait-based social preferences; 2) we use this model to investigate how different trait-based social preferences affect social network structure and function. We find that the preferences can affect the networks' efficiency at transmitting disease and information, and their robustness against fragmentation when individuals disappear, with the effects often – but not always – going in the direction of slower transmission and lower robustness. Furthermore, the extent and form of the effects depend on both the type of preference and the type of trait it is used with. The findings lead to new insights about the potential mechanisms driving the structural diversity of animal social networks, the importance of trait value distributions for social structure, the degree distributions of social networks, and the detectability of trait effects from network data. Overall, the study shows that trait-based social preferences can have effects that go far beyond direct benefits individuals gain from social partner selection, and that the types of preferences which are present in a population can have far-reaching consequences for the population. We discuss the implications of the results for social evolution and the empirical study of animal social networks.

## Lay summary

Animals living in social groups often preferentially socialize with others that have specific characteristics, such as large body size or a specific sex. Using a new computer model, we show how such preferences affect the structure and function of social networks. We find that the preferences can affect disease and information transmission, and the robustness against breakdown of the social structure when individuals are lost. The preferences can therefore have far-reaching consequences for animal groups.

# Introduction

Sociality is a central aspect of life on earth. Across animal species, individuals are embedded in social network structures, which emerge from the patterns of social interactions between individuals. These structures have major consequences on both short and long (evolutionary) timescales. The patterns of social connections affect spreading processes such as the transmission of disease and information (Pastor-Satorras et al. 2015; Evans et al. 2020), and the social connectedness of individuals has been linked to their health, reproduction and survival in multiple species (Snyder-Mackler et al. 2020). Similarly to ecological environments, social structures exert selection pressures and hence act as an evolutionary force (Kurvers et al. 2014; Fisher & McAdam 2017). Understanding social structuring is therefore important for understanding the conditions of life for individuals, and the evolution of social systems and species.

A key question for understanding social structures is how they arise. To gain a comprehensive understanding of social structuring, we must understand in depth how the diverse social network structures observed across different species emerge from individual behavior (Cantor et al. 2021a). More precisely, we need to unravel how different generative processes (such as strategies for social partner choice, and movement patterns based on resource distributions and dispersal) shape different structural and functional aspects of social networks. This is important for our fundamental understanding of sociality, as well as for predicting how changes in social strategies (for example caused by environmental disruptions) may affect social systems (Fisher et al. 2021).

Generative network models are a key tool for investigating how network structures emerge from underlying processes (reviewed in Brask et al. 2025a). These models are in essence algorithms that create simulated networks, based on rules that define how network nodes link to each other. They constitute a central pillar in network science, where they have been used extensively to investigate diverse aspects of networks (for common models see e.g. Erdős & Rényi 1959; Watts & Strogatz 1998; Barabási & Albert 1999; Lee & Wilkinson 2019). Generative network models provide a natural approach for investigating the connection between individual behavior and social network structure, and models based on empirical knowledge about animal social structures make it possible to investigate how generative processes of importance in real animal social systems affect network structure and function.

The empirical study of animal social networks has seen a great development in the last two decades, and social networks have now been quantified and analyzed in many species (Krause et al. 2015). This large body of empirical studies makes it possible to draw inferences about which generative processes are of general importance for the emergence of animal social structures. These generative processes can then be investigated with generative network modelling, to understand how they affect network properties. While generative network models have not yet played as central a role in the study of animal social structures as in wider network science, they have been used to investigate various aspects of these networks, including a number of generative processes (e.g. age-dependent social strategies, Siracusa et al. 2023; cooperativeness-based partner choice, Darden et al. 2020; foraging behavior, Ramos-Ferdández et al. 2006, Cantor & Farine 2018; spatial movement, Chimento & Farine 2024; and social inheritance, Ilany & Akçay 2016). However, a general model for a generative process that is of central importance across species (namely *trait-based social preferences*, explained below), has been missing. Furthermore, it has not yet been systematically investigated how key generative processes of animal social networks affect different aspects of network structure and function. We are therefore in need of a map that reveals how different key processes affect different structural and functional network properties. This could significantly advance our understanding of social systems and identify new important directions for the empirical study of sociality.

One generative process that plays a key role in social networks across species is trait-based social preferences. We use the term *trait-based social preferences* (abbreviated: *trait preferences*) to refer to behavior where individuals preferentially socialize with others that have certain trait values – for example a certain sex, body size, or genetic relatedness. Empirical research shows that such preferences play a central role for social network structures across species (details in next section). Mapping the connection between such preferences and structural and functional network features is therefore important for understanding social systems. But we currently do not understand well how different trait-based social preferences affect social network structure, and how such effects may in turn have consequences for network function, such as network transmission efficiency and robustness.

In this study: 1) we develop a general and flexible generative network model that can create networks based on trait-based social preferences; and 2) we use this model to investigate (via computer simulations) how different trait-based social preferences affect social network structure and function. In the following, we first briefly consider empirical evidence for the importance of trait preferences in animal social networks. We then introduce the model. We thereafter present the methods and results of our investigation of how trait preferences affect network structure and function, and we end with a discussion and conclusion.

## Trait preferences in real social systems

The importance of traits for animal social structures can be seen by the fact that trait values are frequently found to be non-randomly distributed in the networks. The observed patterns typically take two forms: 1) social assortment of individuals by trait values, where individuals are more (or less) connected to individuals that are similar to themselves with respect to the trait; and 2) correlations between trait values and individual social connectedness, where individuals that have certain trait values (such as higher values - for example larger body size) are more socially connected (for example have more social links). Such trait patterns can arise from different trait-based social preferences (as will be illustrated in our analysis). We call these two types of patterns *similarity patterns* and *popularity patterns*, respectively (this fits with terms we use in our model, explained in the next section).

In Table 1 we provide a list of studies that have found trait patterns in animal social networks. This shows that trait patterns in social networks are observed across animal taxa, including insects, reptiles, birds, fish, and mammals (they are also commonly observed in human social networks, e.g. McPherson et al. 2001). These widely observed patterns imply that social preferences based on traits play an important role in the formation of social networks across species, because trait-based social preferences give rise to such non-random trait patterns. Trait-based social preferences can therefore be considered a fundamental generative process of social networks. The empirical studies also show that the preferences that are present in the systems take different forms (giving rise to different observed trait patterns), and that preferences for several traits often act simultaneously in any given social system (as trait patterns are often found for several traits in the same network; Table 1).

Trait patterns can arise from both active and passive social preferences. Experimental studies have demonstrated that individuals actively use the traits of others in their decisions of who to socialize with, and prefer specific others as social partners depending on their traits (i.e. active social preferences, e.g. Guttridge et al. 2009; Jones et al. 2010; Cattelan & Griggio 2020). Non-random socialization based on traits can also arise from an underlying factor instead of active partner choice (i.e. passive social preferences); for example, individuals with similar traits may socialize more with each other due to similar habitat preferences. In this study, we use the term *preference* in a general sense that covers both active and passive preferences, and our model can be used to model both active and passive preferences.

**Table 1. Examples of species for which trait patterns have been found in their social networks.** The list is not exhaustive. The patterns may arise from both active and passive preferences (see main text). *Similarity patterns* refers to social assortment by trait values, and *popularity patterns* refers to correlations between trait values and individual social connectedness (note; assortment and correlations can be both positive and negative). A tick mark indicates that the pattern was found for at least one of the traits investigated in the study. Parenthesis indicates that several networks from different time periods were tested with the same trait(s) and the pattern was found in one/some but not all of the networks. A blank cell indicates that either the pattern was not investigated, or it was investigated but was not found (we do not distinguish between these cases because in both cases the pattern could still exist in the species – for the latter case in other traits than those investigated). *Similarity pattern traits* and *popularity pattern traits* refer to the traits for which similarity and popularity patterns, respectively, were found.

| **Species** | **Similarity patterns** | **Popularity patterns** | **Similarity pattern traits** | **Popularity pattern traits** | **References** |
|---|---|---|---|---|---|
| forked fungus beetle | | ✓ | | sex | Costello et al. 2022 |
| paper wasp | | ✓ | | personality | Laub et al. 2024 |
| sleepy lizard | ✓ | | relatedness, sex | | Leu et al. 2010, Godfrey et al. 2014 |
| great tit | ✓ | ✓ | personality | personality | Aplin et al. 2013, Snijders et al. 2014 |
| Gouldian finch | ✓ | | relatedness, age, head colour, | | Kohn et al. 2022 |
| long-tailed manakin | (✓) | ✓ | social status | social status | Edelman & McDonald 2014 |
| sulphur-crested cockatoo | ✓ | | relatedness, sex | | Penndorf et al. 2023 |
| barnacle goose | ✓ | | relatedness | | Kurvers et al. 2013 |
| guppy | ✓ | ✓ | body size, sex, behavioural score | cooperativeness, behavioural score | Croft et al. 2009, Brask et al. 2019 |
| reef manta ray | ✓ | ✓ | sex, maturity | maturity, reproductive status | Perryman et al. 2019 |
| white shark | ✓ | | body size, sex | | Findlay et al. 2016 |
| sand tiger shark | ✓ | | maturity | | Roose et al. 2022 |
| bottlenose dolphin | ✓ | ✓ | relatedness, age | foraging type | Diaz-Aguirre et al. 2018, Gerber et al. 2020, Foroughirad et al. 2023 |
| humpback dolphin | ✓ | | sex | | Hunt et al. 2019 |
| killer whale | ✓ | ✓ | relatedness, sex, age, diet type | sex, age | Reisinger et al. 2017, Weiss et al. 2021, Jourdain et al. 2024 |
| sea lion | ✓ | ✓ | sex, age | sex, age | Wolf et al. 2007 |
| Natterer's bat | ✓ | ✓ | relatedness, sex, age, breeding status | sex, age, breeding status | Zeus et al. 2018 |

| greater horseshoe bat | ✓ | ✓ | sex, age | sex, age | Finch et al. 2022 |
|---|---|---|---|---|---|
| black-tailed prairie dog | ✓ | ✓ | sex, age, body condition | breeding status | Kusch & Lane 2021a, Kusch & Lane 2021b |
| yellow-bellied marmot | (✓) | ✓ | relatedness, sex, age | age, docility | Wey & Blumstein 2010, Williams et al. 2023 |
| racoon | (✓) | | sex, age | | Hirsch et al. 2013 |
| eastern grey kangaroo | ✓ | | relatedness | | Best et al. 2014 |
| big horn sheep | (✓) | ✓ | relatedness, reproductive status | age, reproductive status | Vander Wal et al. 2016 |
| mountain goat | ✓ | | relatedness, reproductive status | | Godde et al. 2015 |
| giraffe | ✓ | ✓ | relatedness | sex, age | Carter at al. 2013, Ferres et al. 2021 |
| Geoffroy's spider monkey | ✓ | ✓ | relatedness, sex | matriline | Jasso-del Toro et al. 2024 |
| sooty mangabey | ✓ | | relatedness, sex, age, reproductive status, rank | | Mielke et al. 2020 |
| chimpanzee | ✓ | | relatedness, age, rank | | Bray & Gilby 2020 |
| Guinea baboon | ✓ | | relatedness | | Dal Pesco et al. 2021 |
| Assamese macaque | ✓ | | relatedness | | De Moor et al. 2020a, De Moor et al. 2020b |
| rhesus macaque | | ✓ | | sex, age, rank | Brent et al. 2017, Beisner et al. 2020 |

## The trait preference model

Our model, which we call the *trait preference model*, is a general method for generating simulated networks based on trait preferences. Here we provide a verbal description of the model; a mathematical description can be found in Supplementary Material A. An R package with tools for network generation using the model is described in Brask et al. (2025b).

The model works by determining the *social attraction* between each pair of individuals, where a higher social attraction gives a higher chance of the pair being socially connected (having a direct link in the network) and a stronger social connection (a stronger link). The social attraction can be affected by the trait values of the individuals, via trait-based social preferences. As in the real world, multiple traits may simultaneously affect the social attraction. How a given trait affects the social attraction depends on the type of preference it is combined with (preference types are described in the next section). The traits can be of any type (e.g. sex, body size, etc.), and they may each be used with any preference type. For example, for a given modelled network, the social attraction values (and hence the network structure) may be affected simultaneously by social preferences for one's own sex and for large body size.

The importance of each trait (and thereby also the importance of the preference types they are combined with) for the social attraction may differ. This reflects that individuals in the real world may place more weight on one trait than on another when deciding who to socialize with. The traits may also be more or less important overall, reflecting that real networks may be more or less strongly driven by trait preferences. Together this means that the model can generate networks that range from randomly structured to strongly driven by trait preferences, and it can generate networks which are affected by different traits - and different types of preference - in any ratio.

In summary, the social attraction (and thereby the chance and strength of a social connection) between individuals in the modelled population are determined by three things (which are all set by the user of the model): the *trait types* that are included in the model, the *preference types* that each trait is combined with, and the *importance* of each preference-trait combination.

To use the model, we need to provide it with: a) one or more traits (trait value distributions), b) the type of preference used with each trait, and c) the importance of each preference-trait combination. From these factors, a social attraction value for each pair of individuals is calculated. A network is then constructed by a link drawing procedure where a pair of individuals' chance of getting a link (and the link weight in the case of weighted networks) is proportional to the social attraction (for an example, see the procedure of the simulation study, described in the Methods section).

### *Preference types and preference functions*

As a framework to study and model trait-based social preferences, we categorize them into two general *preference types*, which we call *similarity preferences* and *popularity preferences*. They are described in Box 1. The terms *similarity* and *popularity* have been used before with various related meanings in connection with networks (e.g. Papadopoulos et al. 2012); we adapt them to use

**Box 1. Two general types of trait-based social preference**

We categorize trait-based social preferences into the following two general types:

**Similarity preferences:**

Individuals' preference for others as social partners depends on how similar they are to themselves in terms of the trait. They may prefer more similar individuals (known as *homophily* (McPherson et al. 2001) or *assortative mixing* (Newman 2003; Cantwell & Newman 2019)), or less similar individuals (known as *heterophily or disassortative mixing*). For example, individuals may prefer to socialize with individuals of their own sex, or with those that are genetically related (e.g. De Moor et al. 2020a). Here, the individual's preferred trait values depend on its own value of the trait.

**Popularity preferences:**

Certain trait values (such as higher values or lower values) are generally preferred, i.e. popular. For example, individuals may prefer to socialize with those that have a large body size (e.g. Jones et al. 2010), or individuals of one sex may be more popular as social partners than those of the other sex (e.g. Weiss et al. 2021). Here, the individual's preferred trait values do not depend on its own value of the trait.

specifically for trait preferences. The two preference types give rise to the two commonly observed trait patterns described in the previous section (a validation of this is given in the Results section).

We model similarity and popularity preferences using what we call *preference functions*: equations which translate the two trait values of a given trait for a pair of individuals into a trait-specific social attraction. In this way, we can model different types of preference (by changing the function) and combine it with different types of trait (by using different distributions of trait values). Similarity and popularity preferences correspond to different classes of functions (see Suppl. Mat. A and C for equations and details).

We get the overall social attraction between a pair of individuals by combining the trait-specific preference functions (for the number of traits included in the model), taking into account the importance set for each trait (see Suppl. Mat. A for the equation). We note that active and passive preferences (see the previous section) can be expressed via identical preference functions and may be included in the same model.

### *Using the model as a tool for statistical analysis of empirical network data*

Generative network modelling combined with statistical inference is a key approach for understanding observed network structures (Peel et al. 2022). While our main use of the trait preference model in this paper is to generate simulated networks for the investigation of how social preferences affect network structure and function, we here briefly consider the model's potential to be used as a tool for analyzing observed networks. When generating simulated networks without empirical data, the model parameters (the importance of each trait-preference combination) are set by the user. However, the parameters can also be inferred from observed social networks and trait data. This means that the model has the potential to be used for: 1) inferring the importance of different preference-trait combinations for a given observed network structure, and 2) generating simulated network structures based on a given observed network. Estimation of model parameters from data is not our focus in this study, but we provide an example in Suppl. Mat. B of parameter estimation and inference with a simulated 'observed' network to demonstrate the idea. Another way to use the model with empirical data is with user-set parameters but real trait data, which could for example be useful for investigating how a given real trait distribution may affect network structure via different social preferences (see Brask et al. 2025b).

# Methods

We use the trait preference model to study how trait-based social preferences affect social network structure and function. We provide a verbal description of the methods here; mathematical method descriptions are given in Suppl. Mat. C. All simulations described in Methods were done in R (v. 4.2.1; R Core Team. 2022). The simulations described in Suppl Mat. B (mentioned in the previous section) were done in Julia (details in Suppl. Mat. B).

## *General approach*

We study how trait preferences affect social network structure, transmission and robustness. We investigate the effect of the two general preference types (similarity and popularity, Box 1) in combination with different traits. We systematically vary the importance of each preference type from no importance to maximal importance, and generate large sets (ensembles) of networks at regular points across this range. We then quantify the structure, transmission efficiency and robustness of the generated networks. This approach not only shows the effect of the preferences on structural and functional network properties, it also gives information on how the effects change with increasing importance of the preference types (which is important because we have no a priori reason to assume that the effects are linear across the importance range). Our main focus is on the separate effect of the different preferences, but we also study the effect of the preferences when they act simultaneously.

## *Generating networks based on trait preferences*

### *Model specification*

The trait preference model can contain any number of trait preferences. For our simulations, we specify a version of the model that contains two trait preferences: one for each of the two general preference types (similarity and popularity, Box 1). The two preferences may act separately (when the importance of one of them is set to zero) or simultaneously. This model version thus lets us investigate the effect of the two preference types on network structure and function. A mathematical specification of the model version is given in Suppl. Mat. C.

### *Preference types*

We study the two main preference types: popularity and similarity (Box 1). We model the preferences with preference functions that correspond to trait patterns that are well-known from empirical animal social networks (Table 1) and simulate the following behavior: for popularity, individuals prefer those with *higher* trait values as social partners; for similarity, individuals prefer those with trait values that are *closer* to their own value (i.e. those that are more similar to themselves). See the model specification in Suppl. Mat. C for mathematical description of the preference functions.

### *Trait types*

We use three types of trait distributions, which correspond to three traits that are known to be important in animal social networks (see Table 1 for references, and see Suppl. Mat. C for mathematical specifications of the distributions):

***'Categorical', corresponding to sex:*** Trait values are categorical with two categories (two sexes).

***'Continuous normal', corresponding to body size:*** Trait values are continuous and are drawn from a normal distribution.

***'Continuous circular', corresponding to relatedness:*** Trait values are continuous and are drawn from a circular scale. The fact that the scale is circular means that the closeness of trait values on the scale can correspond to genetic similarity of the individuals, i.e. relatedness.

We generate simulated trait values for the individuals in the networks by drawing values from the given distributions.

We note that the distributions could also be interpreted as other traits that follow the given distributions of trait values (and the results also hold for such traits); therefore, for clarity we refer to the traits not only by their trait name but also by their distribution (as above). In the following, for convenience we furthermore refer to each of the above traits as the 'similarity trait' when it is combined with the similarity preference, and the 'popularity trait' when it is combined with the popularity preference.

### *Network construction*

To generate networks, we use the above-described model version with two traits (the 'popularity trait' and the 'similarity trait'). We use all pairings of the three traits, except that the circular trait is only used as a similarity trait (where similarity of the trait values corresponds to relatedness). For each pair of traits, we vary the importance of each preference type from no importance (where the trait preference has no influence on social attraction and the network structure is random) to maximum importance (where the trait preference has high influence on social attraction and thus could potentially have strong effects on the network), and generate network ensembles across the range of relative importance of the two preference types (see Suppl. Mat. C for further information about the importance values). This gives us network ensembles both for the cases where either preference type is acting alone (when the other has no importance), and when they are acting together. Each ensemble consists of 100 networks.

To generate a network, we use the following procedure: We first draw trait values from the relevant trait distributions for the two included trait types, and assign them to the individuals. We then calculate the social attraction between each pair of individuals, using the model equations of the specified model (Suppl. Mat. C). We then draw social links (network edges) and link strengths (edge weights). The number of links to be drawn is given by a set average degree (average number of links per individual), and the links are drawn by a weighted draw, where the chance of a given link to be drawn, and its link strength, are relative to the social attraction of the two individuals that would be connected by that link (see Suppl. Mat. C for further details about the link drawing procedure). We use a network size of 100 nodes (individuals) and an average degree of 10 to reflect real-world social networks, which are of such relatively small sizes for many species (we checked that this choice did not qualitatively affect conclusions). We use only networks with a single component (that is, all nodes are at least indirectly connected), such that each network corresponds to a single, unfragmented population.

### *Measuring structural and functional properties of the generated networks*

#### *Quantification of social network structure*

To quantify network structure, we measure the following six global network measures, which are averaged for each network ensemble (descriptions of their calculation are provided in Suppl. Mat. C):

To investigate the effect of trait preferences on social network structure, we use the following four global network metrics, which measure different aspects of network structure. We calculate both unweighted metric versions (based on the presence/absence of edges) and weighted versions (based on edge weights). *Unweighted degree* is the number of social links (edges) an individual (node) has, and *weighted degree* is the sum of its edge weights.

***Degree variation:*** The variation in social connectedness across all individuals.

***Degree assortativity:*** The extent to which individuals are primarily connected to others with a similar level of social connectedness.

***Clustering:*** The extent to which individuals' social partners are connected to each other (averaged over all individuals).

***Mean distance:*** The average social distance (shortest path through the network) between individuals.

We furthermore quantify two additional network measures, the purpose of which is primarily to investigate whether the two preference types (similarity and popularity) as expected induce the two common trait patterns in the networks considered in Table 1. When individuals use popularity preferences with a trait, this should (everything else equal) induce correlation between individuals' values of that trait and their degrees (as individuals with the more popular trait values get more links), and when individuals use similarity preferences with a trait, this should induce assortment by that trait (as individuals with (dis)similar trait values prefer each other). We therefore quantify the following measures:

***Popularity trait – degree correlation:*** the correlation between the individuals' popularity trait values and their numbers of social links (unweighted degree).

***Similarity trait assortativity:*** The extent to which individuals are connected to others with similar similarity trait values.

### *Quantification of social network transmission efficiency*

To investigate how trait preferences affect social network transmission, we simulate transmission processes in all the generated networks of each ensemble, and quantify the transmission efficiency of the networks (measured by how quickly the transmission spreads to a given percentage of the population). Each simulation begins with a single, randomly picked infected node, and proceeds in discrete timesteps. For our main analysis, we measure transmission efficiency as the number of timesteps it takes until all nodes are infected. We also explore the transmission efficiency at other points during the transmission process (when 25%, 50% and 75% of the individuals have been infected).

We study two types of transmission: simple and complex. In the former, the probability of infection is usually relative to the *number* of infected neighbours, whereas in the latter, the probability of infection depends on the infection status of others in different ways, and is often related to the *proportion* of infected neighbours (Firth 2020). The transmission models we use are described below, and mathematical descriptions of them are given in Suppl. Mat. C. For each transmission type, we study both the case where edge weights are affecting the transmission, and the case where they are not, giving us four transmission subtypes. For each network, we run 10 replications of a given transmission subtype (giving 1000 simulations for each transmission type for a given network ensemble). We calculate the transmission efficiency of the network as the inverse average transmission time.

***Simple transmission***: We use the SI model to model simple transmission. Here, the probability of being infected depends on the number of infected network neighbors, or the social connectivity (summed edge weights) that goes to the infected neighbours (depending on whether edge weights affect the transmission), and infected individuals stay infected. This is a relevant model for the spread of disease and information.

***Complex transmission***: We use a proportional transmission model to model complex transmission (Centola & Macy 2007; Guilbeault et al. 2018; Evans et al. 2020; Firth 2020). Here, the probability of being infected is relative to the proportion of network neighbors that are infected, or the proportion of social connectivity (proportion of summed edge weights) that goes to infected neighbors (depending on whether edge weights affect the transmission), and infected individuals stay infected. This is a relevant model for the spread of information and behaviors.

### *Quantification of social network robustness*

To investigate how trait preferences affect social network robustness, we study the breakdown of the network structures when individuals are lost from the population. We simulate the loss of individuals in each of the generated networks by sequentially removing nodes, until two nodes are left. We measure robustness as the time at which the original single network component first breaks down, i.e. the number of individuals that can be lost before the network breaks into pieces. As we do not know a priori if different aspects of robustness are affected differently, we also study three other robustness dimensions (described in Suppl. Mat. C).

We focus on four types of loss of individuals, where the individuals (network nodes) are removed either in random order or in order according to measures of their social connectedness (node centrality metrics):

***Random loss:*** nodes are removed in random order. Corresponding to a situation where the death or dispersal of individuals is independent of their social network position.

***Loss based on social connectedness*****:** nodes are removed in order corresponding to their degrees (unweighted or weighted), with highest degree first. Corresponding to a situation where individuals with high social connectedness are more likely to die or disperse (for example trophy hunting of older hub individuals).

***Loss based on social isolation***: nodes are removed in order corresponding to their degrees (unweighted or weighted), with lowest degree first. Corresponding to a situation where individuals with low social connectedness are more likely to die or disperse (for example when socially isolated individuals have low health).

***Loss based on social bridging*****:** nodes are removed in order corresponding to their betweenness, with highest betweenness first. Corresponding to a situation where individuals that connect social subgroups are more likely to die or disperse (for example if individuals moving between social subgroups have an increased risk of dying).

## Results

### *Effects of trait preferences on social network structure*

The results confirm that the two preference types give rise to the expected trait patterns, with popularity preferences leading to correlations between individual social connectedness (degree) and trait values, and similarity preferences leading to social assortment based on trait values (Fig. 1B; see Table 1 for empirical evidence of the trait patterns).

Looking at the effect of trait-based social preferences on network metrics, we see that the different preferences affect network structure differently and that both the preference type, and the trait type it is combined with, are important for the effect (Fig. 1C). Thus, it can make a difference for a population's social structure which preference types are present in the population, and with which traits they are used.

We found similar results to the one in Fig. 1 for larger networks and other average degrees, and for versions of the network metrics that take edge weights into account (Supplementary figures S1, S2 and S3).

We see that clustering (the extent to which individuals' neighbors are connected to each other) and the social distance between individuals are strongest affected by similarity preferences, whereas degree variation (the variation in the number of links individuals have) and degree assortativity (the extent to which individuals connect to others of similar connectedness) are strongest affected by popularity preferences. All the investigated preference-trait combinations have effects on network structure, but on different scales (Fig. 1; see Fig. S4 for plots with the continuous traits only, for a more detailed visualization of their effects). In particular, we often see stronger effects when the preferences are combined with a categorical trait.

We focused on a categorical trait with two categories, as this corresponds to a key trait of importance for social networks in many species (sex), but more than two trait categories may also occur in real-world systems (e.g. multiple matrilines, De Moor et al. 2020a) and we find some variation in effect depending on the number of categories (Fig. S5).

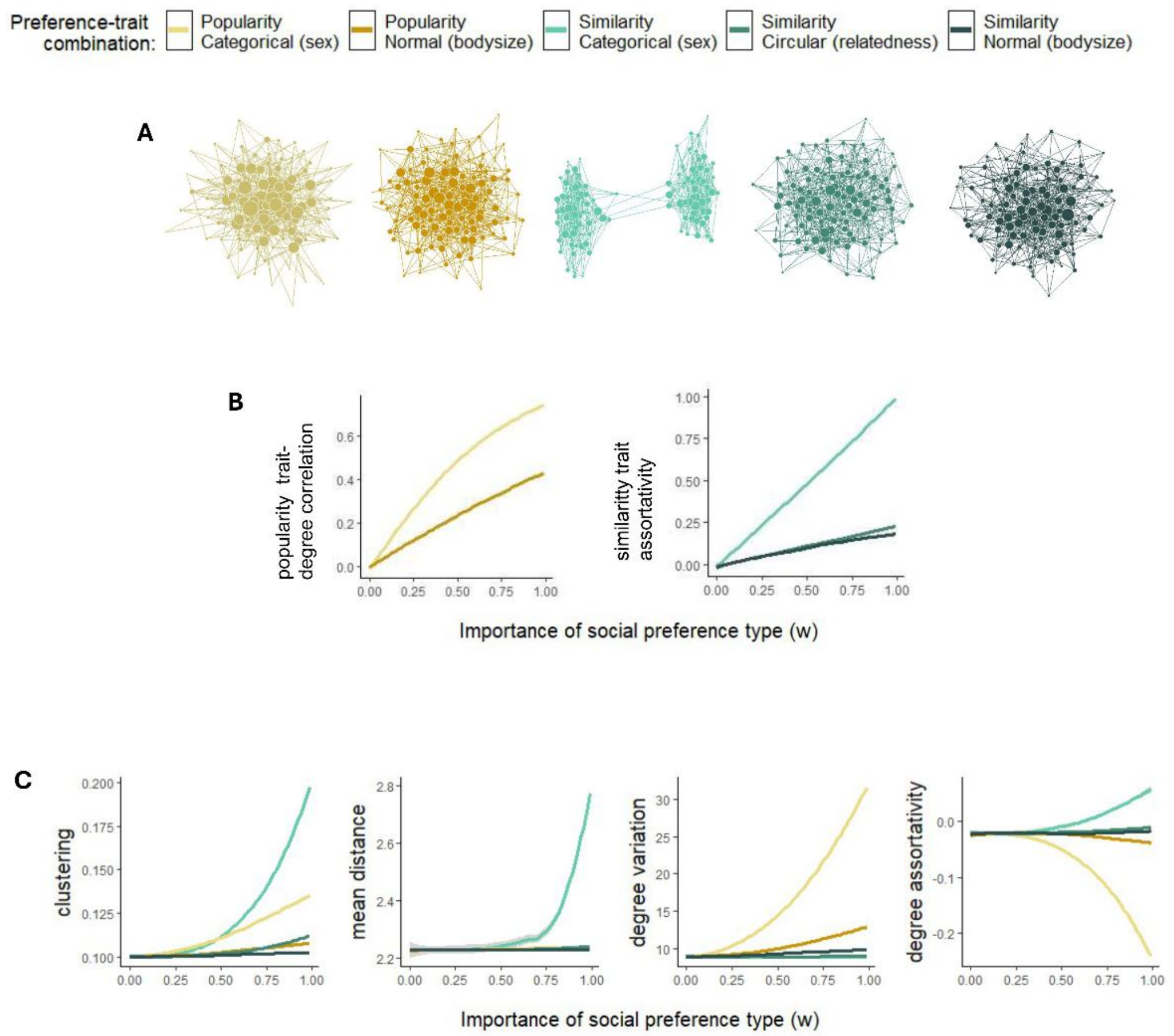

**Fig. 1. Effects of trait-based social preferences on social network structure.** Results for each combination of preference type (popularity, similarity) and trait type are shown with the same color throughout the figure (see legend). (A) Example networks for each preference-trait combination (made with maximal importance of the social preferences). Node sizes correspond to degree. (B) Each of the two preference types gives rise to a common trait pattern. Results are shown for each preference-trait combination. (C) The change in four structural network properties (global network metrics) with increasing importance of social preferences, for each preference-trait combination.

Another way to consider the effects of the preferences is to look at how they move networks around in structural spaces (Fig. 2). Looking at the categorical trait (which gives the strongest effects for both preference types), we see that similarity preferences have particularly strong influence on the position of the networks in a structural space that we name the small-world space (consisting of two structural dimensions of importance in small-world networks; Watts & Strogatz 1998), whereas popularity preferences has particularly strong influence on their position in the space that we name the degree space (consisting of two degree-related dimensions). This exemplifies how the trait-based social preferences used in a population can influence - and limit - the areas of structural space that the social network of the population can occupy.

Finally, looking at the networks' degree distributions (i.e. the distribution of the number of links connected to each individual), we find that the different preference-trait combinations all lead to degree distributions that are largely symmetric, rather than skewed or long-tailed (Fig. S6).

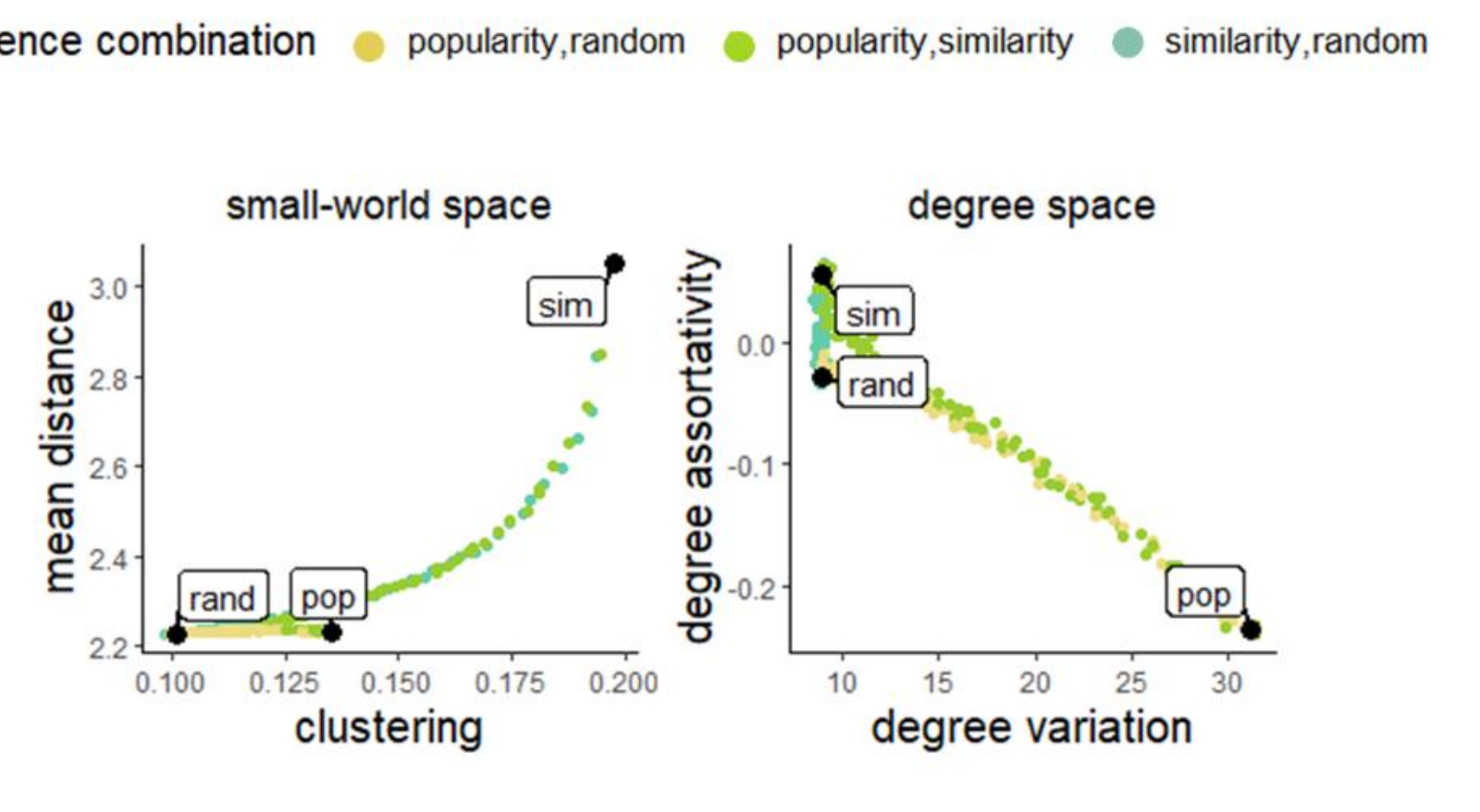

**Fig. 2. Movement of networks in structural spaces caused by trait-based social preferences**. Shown for the categorical trait (sex). Each dot indicates the average of 100 networks. Black dots indicate positions of networks that are generated based on one mechanism with maximal importance (randomness, popularity, or similarity). Colored dots indicate networks based on a combination of two mechanisms (see legend) across the range of relative importance of the two mechanisms.

### *Effects of trait preferences on social network function*

We find that the effects of trait-based social preferences on social structure can lead to changes in social network transmission and robustness, and the extent of these changes depends on the preference-trait combination, as well as on the type of transmission and the type of loss of individuals (Fig. 3). Thus, which types of preferences are used in a population, and which types of traits they are used with, can have consequences for the vulnerability of the populations to disease, their efficiency of cultural transmission, and their resilience to structural breakdown when individuals disappear. In the following, we describe the general findings concerning transmission and robustness; detailed explanations of mechanisms behind various aspects of these results can be found in Suppl. Mat. D.

#### *Effects on social network transmission efficiency*

Results for transmission processes that do not depend on edge weights (Fig. 3A) and that do depend on edge weights (Fig. S7) are quite similar. There are some differences in the effect of the preferences on the two types of transmission (Fig. 3A and S7).

Overall, we see that social preferences in most cases give rise to networks that have slower transmission (Fig. 3A and S7), although similarity preferences in particular also in some cases increase transmission (Fig. S7). Thus, both types of preference may protect against the spread of disease and slow the spread of information and behaviors, and similarity preferences may also do the opposite under some conditions.

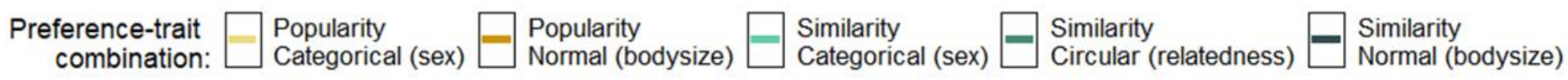

A. Effect of social preferences on social network transmission

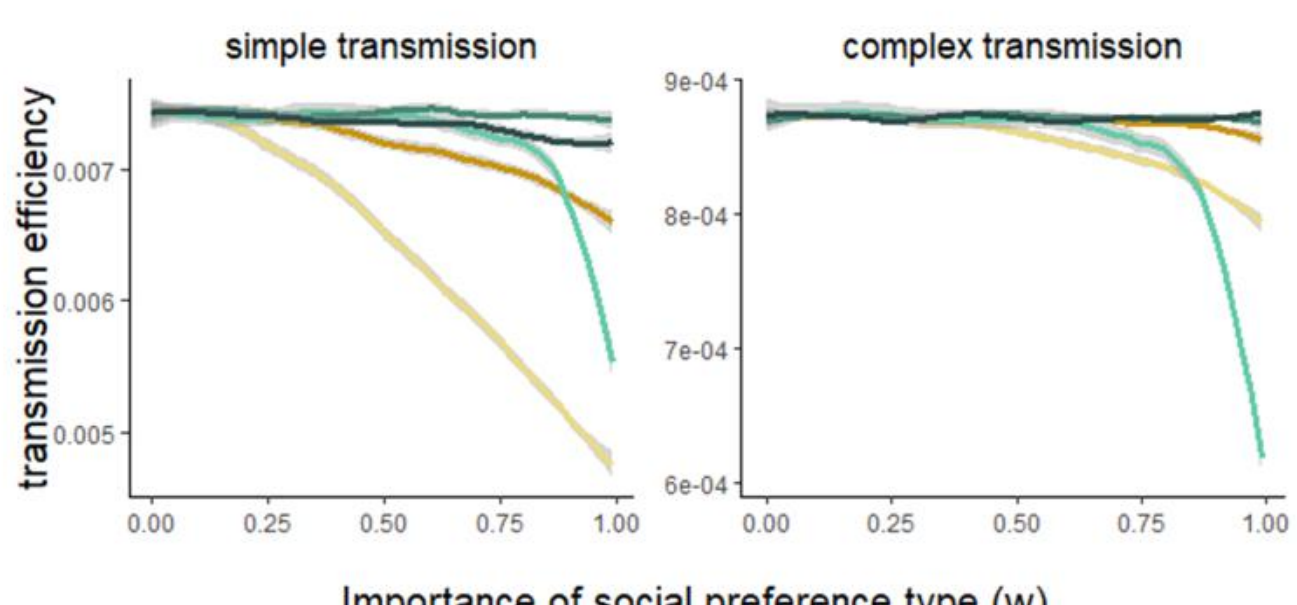

B. Effect of social preferences on social network robustness

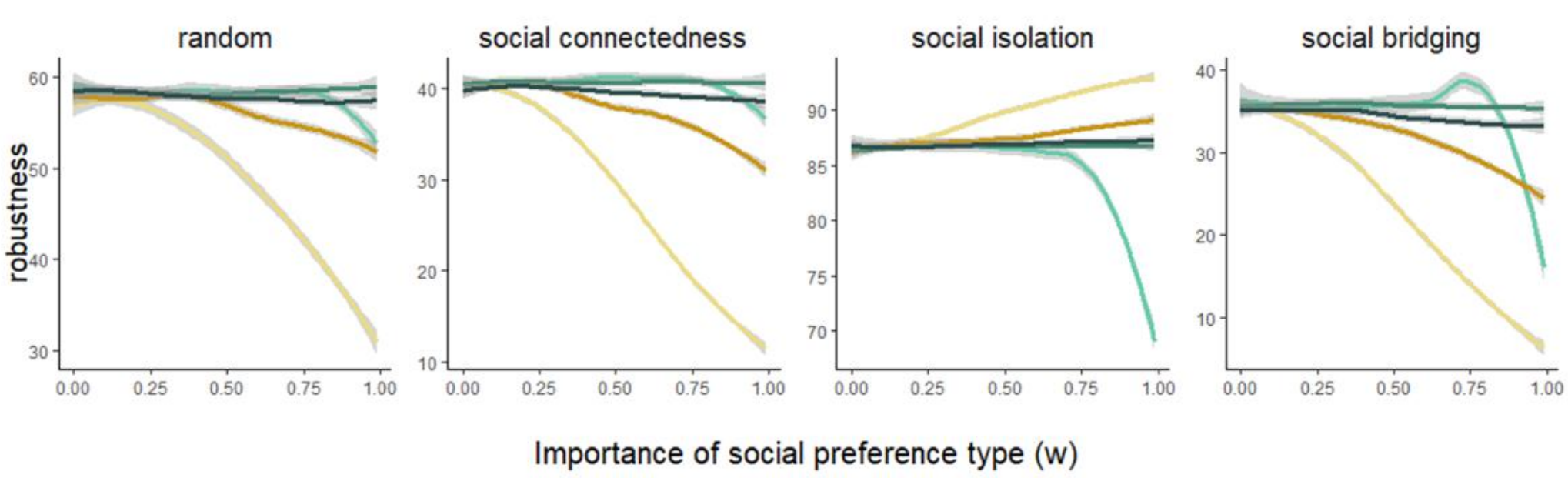

**Fig. 3. Effects of trait-based social preferences on social network transmission and social network robustness.** Results for each combination of preference type (popularity, similarity) and trait type are shown with the same color throughout the figure (see legend). (A) The change in social network transmission efficiency with increasing importance of social preferences, for each preference-trait combination and two types of transmission. (B) The change in social network robustness (component break time) with increasing importance of social preferences, for each preference-trait combination and four types of loss of individuals (network nodes; the loss is either random or based on measures of the individuals' network positions).

The results for lower percentages of infected individuals show that particularly the effects of popularity preferences on simple transmission happen towards the end of the simulations (Fig. S8 versus Fig. 3A) – that is, the popularity preferences mostly affect the ability of simple transmission to reach the most remote nodes. For the case of popularity preferences, the slower transmission is thus likely to be caused by increased centralization and core-periphery structure, where peripheral nodes are harder to reach (particularly for simple transmission); in contrast, for similarity the slower transmission is likely caused by increased modularity (see Suppl. Mat. D for details). The differences in the effect of preferences on the two types of transmission may be explained by the transmission types' differing ability to reach peripheral nodes (see Suppl. Mat. D for details).

### *Effects on social network robustness*

Network robustness (the component break time) where loss of individuals is based on unweighted (Fig. 3B) and weighted (Fig. S9) centrality measures give very similar results. Results for the other investigated robustness dimensions are also mostly very similar (Fig. S10).

Overall, we see that trait preferences can lead to both increased and decreased social network robustness, with the latter being more common. The effect depends on the type of preference and trait, and the type of loss of individuals. The results imply that the social preferences used in populations can have consequences for their risk of fragmenting when individuals disappear – and thereby also for the amount of behavioral compensation (rewiring of social connections) needed to avoid such fragmentation (Shizuka & Johnson 2020).

Most of the clear effects on network robustness are caused by the popularity preferences, and may be explained by these preferences leading to increased degree variation and core-periphery structure (see Suppl. Mat. D for details).

## Discussion

Empirical studies imply that trait-based social preferences is a key generative process for animal social structures. Here we have introduced a generative network model that creates networks based on trait-based social preferences. The model can be used with any number of traits that can simultaneously affect the network, and with different preference types, different trait types, and different relative importance of each trait-preference combination. It can also generate networks across the continuum from random structure to structure strongly driven by preferences. The model thus provides a flexible tool for generating networks from trait-based social preferences.

Using this model, we investigated the effect of trait-based social preferences on social network structure and function. The results indicate that the preferences can have consequences for the vulnerability of populations to epidemics, their efficiency in spreading cultural information, and their ability to withstand breakdown of their social structure when individuals are lost from the population. The extent and direction of the effects depend both on the type of preference and the type of trait, and the preferences can affect the populations both positively and negatively. Thus, the trait preferences used in a population can have far-reaching consequences for the population, as they can affect not only the direct benefits that individuals gain from social partners, but also the higher-level function of the social system. The results lead to a number of insights which we consider in the following.

### ***Structural diversity of networks based on trait preferences***

Animal social networks show diverse structures. For example, the structures of some species and populations are characterized by clear division into social clusters (network communities, Javed et al. 2018), while others show more homogeneous or centralized linking patterns. We find that networks generated with the trait preference model show structural diversity (example in Fig. 4), with structural features varying depending on the trait types, preference types, and the importance of each preference-trait combination. The model was not designed with the aim of creating diverse structures, but with the aim of modelling networks based on different trait preferences. The diversity observed in its output networks (together with the common presence of trait preferences in real animal social networks, Table 1) therefore suggests that differences in trait preferences could potentially be an important explanatory factor for the structural diversity observed in real-world social networks. Trait preferences may thus play an important role in the creation of this diversity – in combination with other relevant factors such as variation in resource distributions and trait-independent movement patterns (e.g. Ramos-Ferdández et al. 2006, Cantor & Farine 2018, Chimento & Farine 2024). Furthermore, structural variation between networks is well-known to lead to differences in functional aspects of the systems, such as their robustness and their efficiency in spreading disease, information and behaviours (e.g. Sah et a. 2018, Evans et al. 2020, Cantor et al. 2021b, Brask & Brask 2024), and the structural diversity of the generated networks (together with our simulations) imply that differences between systems in trait preferences may be an important underlying factor for differences in network function.

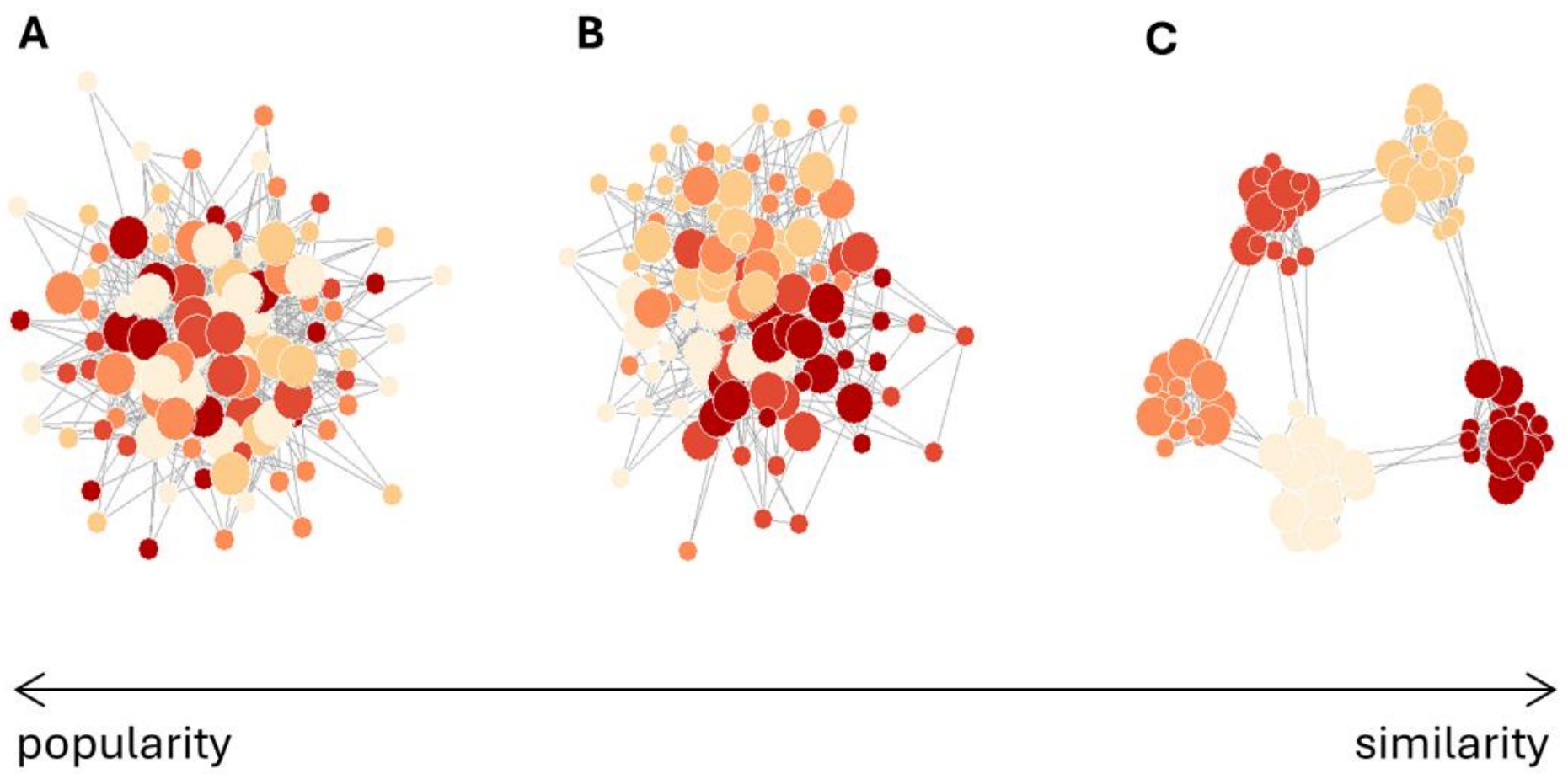

**Fig. 4. An example of structural diversity in networks generated with the trait preference model.** Examples of networks created with a version of the model that includes two traits, which are used with popularity and similarity preferences respectively (as in the simulation study of this paper). Both traits are here categorical. Note, this example does not cover the full diversity of structures that can be generated with the model. Node colours correspond to the categories of the similarity trait (5 categories) and node sizes correspond to the categories of the popularity trait (2 categories). (A) High importance of popularity preferences, no importance of similarity preferences; (B) Equal importance of popularity and similarity preferences; (C) High importance of similarity preferences, no importance of popularity preferences.

### *The role of trait types in trait preference effects*

An important insight from the results is that the effect of trait-based social preferences depends not only on the preference type but also on the type of trait the preference type is used with. This implies that the distributions of trait values in animal populations play a key role for the effect of trait preferences on network structure and function. The study thus indicates that to understand the role of traits in the emergence of real social structures, we must consider not only preference types but also trait value distributions.

The observed effects of trait type have interesting implications for real-world systems. In particular, they suggest that social preferences concerning the sex of social partners may have a disproportionately large effect on network structure and function, compared to other common traits. For example, if individuals have preferences for socializing with others of own sex, this could have a much larger effect on the network than if individuals have preferences for socializing with others of a similar body size (for the same level of preference importance). It is possible, however, that individuals could use preference functions that effectively make continuous traits act like categorical traits (for example if individuals prefer those of a body size larger than average, while not discriminating between individuals within the two categories of below-average and above-average). In that case, traits with continuous distributions could have equally strong effects on the structure and function of networks as categorical traits (for a given level of preference importance). The presence of such discretizing preference functions in real social systems could potentially be investigated with behavioral choice experiments (Guttridge et al. 2009; Jones et al. 2010; Cattelan & Griggio 2020).

### *Effects of trait preferences on degree distributions*

The degree distribution of a network can have consequences for network processes. This network feature has therefore received much attention in wider network science, and it has been concluded that many real-world networks have power-law distributed degrees (where a few nodes have a disproportionately large number of links and many nodes have few links; Barabási & Albert 1999, Broido & Clauset 2019, Holme 2019). In contrast to other realms of network science, the degree distributions of animal social networks have not been a significant research focus (for an exception see James et al. 2017), and underlying distributions may be harder to deduce from data due to the relatively small size of the observed networks. We found that the degree distributions of networks created with the different preference-trait combinations are all largely symmetric (Fig. S6), rather than following power-laws (i.e. long-tailed). Given the empirical indications of trait preferences being widespread across social networks of different species (Table 1), our results suggest that semi-symmetric degree distributions (rather than power-law distributions) could potentially be frequently underlying real animal social networks.

### *Nonlinearity and detectability of trait preference effects*

The results suggest that effects of trait preferences on network structure and function cannot be expected to increase linearly with the importance of the preferences (Fig. 1C and Fig. 3). Non-linearity is particularly pronounced for similarity preferences with a categorical trait (such as preference for own sex). This implies that populations using this preference may have network structures close to random unless the preferences are of high importance. Similarity preferences with a categorical trait may, however, still be detectable in real networks when they are not highly important, because trait assortment increases linearly with preference importance (Fig. 1B). In fact, All the preference-trait combinations have effects on the trait-based measures that are either linear or concave (Fig. 1B), implying reasonable detectability from these measures. Together, these results furthermore imply that if a given extent of trait assortativity or centrality-trait correlation is found in a real network, it cannot just be assumed that the trait has a proportionally large effect on other aspects of network structure and function (Fig. 1B versus Fig. 1C). Deducing those effects for a given system may benefit from approaches that combine empirical data analysis and network modelling.

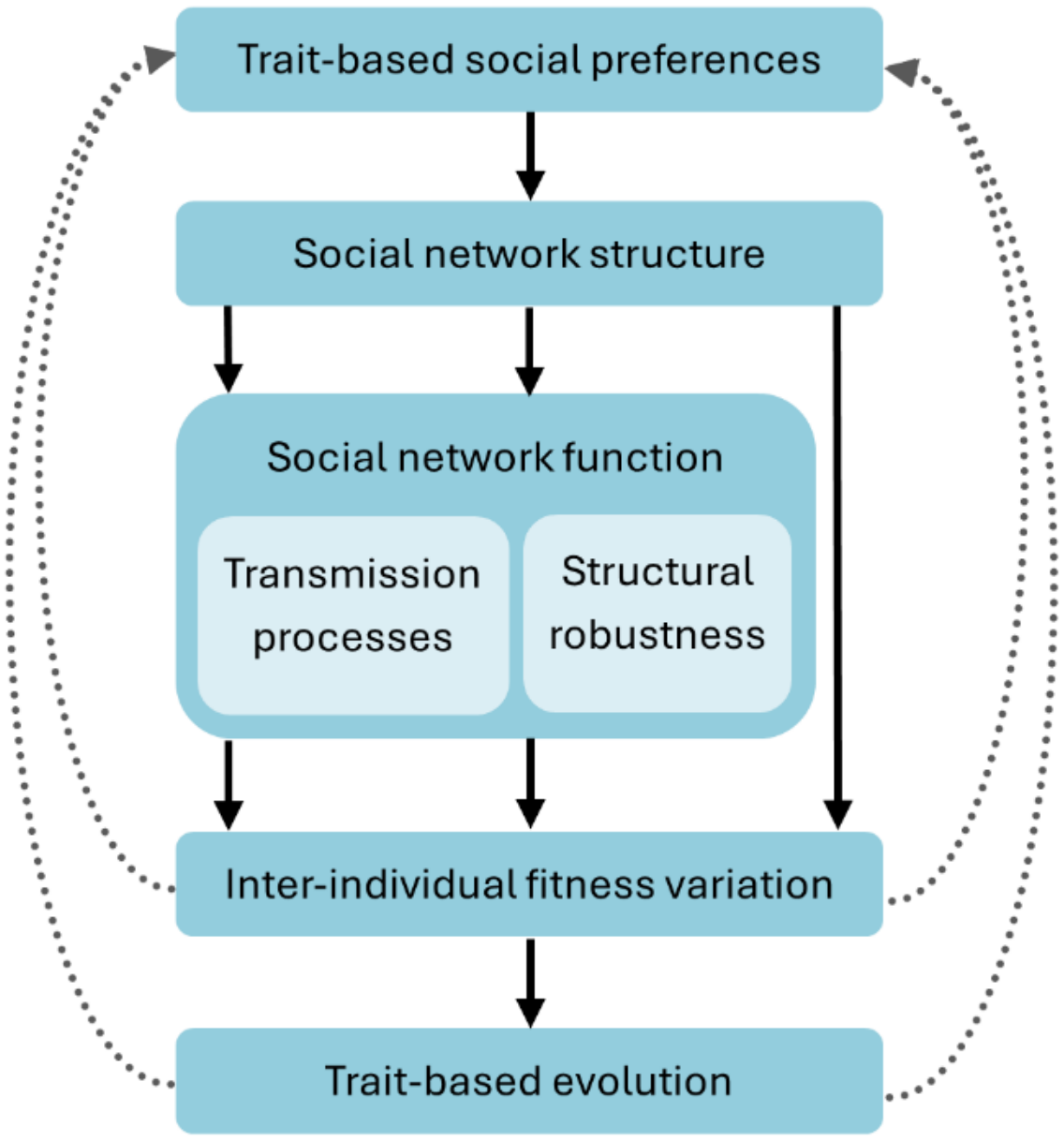

**Fig. 5. Potential effects of trait-based social preferences on evolutionary processes.** The types of trait-based social preference used in a population have consequences for social network structure, and this has potential implications for evolutionary processes. Effects of trait-based social preferences on network structure and function (such as those shown in this study) could lead to changes in relative fitness between individuals. (For example, increased degree variability from popularity preferences could affect fitness variation, when fitness is linked to social connectivity; Snyder-Mackler et al. 2020). Such changes to inter-individual fitness variation could have consequences for the selection on traits and thereby affect their evolution. Changes to fitness and trait distributions could in turn affect trait-based social preferences, leading to a feedback loop.

### *Evolutionary consequences of trait preferences*

An interesting implication of our results concerns the consequences of trait-based social preferences for the evolution of populations and species. The role of social network structure in biological evolution has gained considerable recent interest (Kurvers et al. 2014; Fisher & McAdam 2017; Sueur et al. 2019; Cantor et al. 2021a; Wice & Saltz 2021). While the simulations of this study do not quantify evolutionary effects, the fact that trait-based social preferences were found to affect social network structure and function imply that such preferences can affect the evolutionary trajectory of populations (see Fig. 5). Our results thus suggest that it matters for the evolution of a population (including the evolution of traits and behavior) which trait preferences are used in the system. Investigating the evolutionary implications of different trait-based social preferences provides an exciting area for future research.

## Conclusion

In this study, we have presented a general model for the generation of networks based on trait preferences, and we have used it to investigate the link between trait preferences and network structure and function. The study showed that it can have wide-ranging consequences for populations which trait preferences they use, and the results also lead to a number of further insights that are of relevance for real animal social networks. They implied that trait preferences may be important for explaining the diversity observed in animal social networks; that both preference types and trait types (trait value distributions) are important to consider for understanding the emergence of social structures; that animal social networks may often have symmetric degree distributions (contrary to a general assumption); and that the extent to which traits affect network structure and function cannot be directly deduced from the extent of trait patterns detected in observed networks. The results furthermore suggested that the trait preferences used in a given population may have important consequences for its evolutionary trajectory. The study also presents a structural way of thinking about trait preferences, where they are considered as preference functions that translate trait values into social attraction, which provides a framework for studying them. We hope that this study will inspire to new investigations of trait-based social preferences, and we think that studies combining empirical and modelling-based knowledge and approaches may be particularly important for increasing our understanding of the emergence of animal social structures.

## Supplementary material

Supplementary material A-D and supplementary figures are available at Behavioural Ecology online.

## Code availability

For researchers wanting to generate, visualize and quantify networks based on the trait preference model, we suggest using our R package 'okaapi' (introduced in Brask et al. 2025b), which contains user-friendly tools for this. For documentation for this paper, the code for the simulations is available at https://github.com/bohrbrask/documentation-code-for-trait-prefs-paper

## Acknowledgements

We thank Michael N. Weiss, Samuel Ellis, Lauren J. N. Brent and Benjamin F. Maier for helpful input and discussion.

## Funding

This work was supported by the Carlsberg Foundation (CF20-0663 to JBB) and the Novo Nordisk Foundation (NNF23OC0085517 to JBB).

## Author contributions

Conceptualization: JBB, AK, DPC, SL. Formal analysis: JBB, AK. Funding Acquisition: JBB. Investigation: JBB, AK. Methodology: JBB, AK, DPC, SL. Project administration: JBB. Software: JBB, AK. Supervision: SL, DPC. Visualization: JBB, AK. Writing – original draft: JBB. Writing – review & Editing: JBB, AK, DPC, SL.

## Conflict of interest

We declare no conflict of interest.